\documentclass[useAMS,usenatbib]{mn2e}

\usepackage{tabularx}
\usepackage{multirow}
\usepackage{url}
\usepackage{amsmath}
\usepackage{aas_macros}
\usepackage{graphicx}
\usepackage{subfigure}

\title[MRI and shear]{Sensitivity of the Magnetorotational Instability to the shear parameter in stratified simulations}

\author[F. Nauman and E. G. Blackman]{Farrukh Nauman$^{1}$\thanks{E-mail:
fnauman@pas.rochester.edu} and Eric G. Blackman $^{1,2}$\thanks{E-mail: blackman@pas.rochester.edu} \thanks{Simons Fellow; IBM-Einstein Fellow }\\
$^{1}$Department of Physics and Astronomy, University of Rochester, Rochester, NY 14627, USA\\
$^{2}$School of Natural Sciences, Institute for Advanced Study, Princeton NJ, 08540 USA\\}
\begin{document}

\date{\today}

\pagerange{\pageref{firstpage}--\pageref{lastpage}} \pubyear{2014}

\maketitle

\label{firstpage}

\begin{abstract}
The magnetorotational instability (MRI) is a shear instability and thus its sensitivity to the shear parameter $q = - d\ln\Omega/d\ln r $ is of interest to investigate. Motivated by astrophysical disks, most (but not all) previous MRI studies have focused on the Keplerian value of $ q=1.5$. Using simulation with 8 vertical density scale heights, we contribute to the subset of studies addressing the the effect of varying $q$ in stratified numerical simulations. We discuss why shearing boxes cannot easily be used to study $q>2$ and thus focus on $q<2$. As per previous simulations, which were either unstratified or stratified with a smaller vertical domain, we find that the $q$ dependence of stress for the stratified case is not linear, contrary to the Shakura-Sunyaev model. We find that the scaling agrees with \cite{1996MNRAS.281L..21A} who found it to be proportional to the shear to vorticity ratio $q/(2-q)$. We also find however, that the shape of the magnetic and kinetic energy spectra are relatively insensitive to $q$ and that the ratio of Maxwell stress to magnetic energy ratio also remains nearly independent of $q$. This is consistent with a theoretical argument in which the rate of amplification of the azimuthal field depends linearly on $q$ and the turbulent correlation time $\tau$ depends inversely on $q$. As such, we measure the correlation time of the turbulence and find that indeed it is inversely proportional to $q$.
\end{abstract}


\section{Introduction}
The magnetorotational instability (MRI) (\cite{1991ApJ...376..214B}, \cite{1998RvMP...70....1B}) has emerged as a strong candidate to explain turbulence in Keplerian ($q=- d\ln\Omega/d\ln r=1.5$) accretion discs. Since MRI generated stresses draw energy from the shear, the influence of the shear strength on properties associated with the turbulent stresses and transport are of interest to study. Doing so helps to distill the underlying physical mechanisms of transport and  to better inform mean field theories, even if the primary application is ultimately $q=3/2$. Previous  analytical and numerical work has shown that the turbulent stresses are sensitive to $q$ (e.g. \cite{1996MNRAS.281L..21A}, \cite{1998bhad.conf.....K}, \cite{1999ApJ...518..394H}, \cite{2001A&A...378..668Z}, \cite{2003MNRAS.340..969O}, \cite{2006MNRAS.372..183P}, \cite{2008MNRAS.383..683P}, \cite{2009AN....330...92L}) and differ in their sensitivity from that predicted by a Shakura-Sunyaev paradigm.

Here we focus on the $q$-dependence of the MRI for stratified isothermal shearing box simulations. While stratified simulations for different shear values have been studied by \cite{1996MNRAS.281L..21A} and \cite{2001A&A...378..668Z}, their use of vertical domains of 2 density scale heights from the midplane means that their entire box was unstable to MRI, and buoyancy did not play a significant role. Larger vertical domain sizes show that the plasma beta $\beta = 2 \mu_0 p/B^2$, where p is the pressure and B is the magnetic field, decreases below unity at around 2 scale heights away from the midplane (e.g., \cite{2014MNRAS.441.1855N}). We call the region above this point the corona, and the MRI unstable region the disc.

As noted by \cite{1996MNRAS.281L..21A} and \cite{2013MNRAS.428.2255P}, the MRI dependence on the shear parameter is of interest for accretion flows around black holes. Using global general relativistic simulations of accretion flows around rotating black holes, \cite{2013MNRAS.428.2255P} showed that the shear parameter rises close to 2 near the ISCO and then quickly falls down to sub-Keplerian values closer to the black hole.

MRI shearing box simulations suffer from several limitations (e.g. \cite{2007MNRAS.376.1740K}, \cite{2008A&A...481...21R}, \cite{2011ApJ...738...84H}, \cite{2014MNRAS.441.1855N}, \cite{2014arXiv1405.3991H}). We do not study numerical or physical limitations of the shearing box model in this paper but it is important to keep these limitations in mind when interpreting the results of any shearing box simulation.

We assess the dependence of MRI generated turbulence for different values of the shear parameter. We focus particularly on the saturated stresses in the disc region as well as the magnetic and kinetic energy. We describe our numerical setup in section 2. In section 3, we discuss shear dependence of various physical quantities. We conclude in section 4.

\section{Numerical methods}

\subsection{Parameters and Setup of Runs} 
We make use of the finite volume high order Godunov code ATHENA (\cite{2005JCoPh.205..509G}, \cite{2008ApJS..178..137S}). The setup is similar to \cite{2014MNRAS.441.1855N}: the ideal MHD equations are solved in the shearing box approximation with an isothermal equation of state $p=\rho c_s^2$, where $p = \text{pressure}, \rho = \text{density}, c_s = \text{sound speed}$. $\mu_0=1$ in code units. The orbital advection algorithm \citep{2010ApJS..189..142S} is used to speed up the simulation. Our simulations include gravity with an equilibrium density profile $\rho = \rho_0 \exp (-z^2/H^2) $ where $H = \sqrt{2} c_s/\Omega = \text{scale height}$ and $\Omega = \text{angular velocity}$. For initial conditions, we use $\Omega = 10^{-3}, \rho_0 = 1, p_0 = 5 \times 10^{-7}$ and a constant $\beta = 2 p_0/B_0^2 = 1600$, which gives the magnetic pressure the same initial vertical profile as the density.

\begin{table}
\centering
\begin{tabular}{| c | c | c | c |}
\hline \hline
  Shear & Orbits & $\alpha$ & $\alpha_{\text{mag}}$ \\ 
  $-d\ln\Omega/d\ln r$ & $(2\pi / \Omega)$ & $(\times 10^{-3})$ & $(\times 10^{-1})$\\ \hline \hline
  0.7 & 278 & $0.8770 \pm 0.2803$ & $2.481 \pm 0.1625$ \\
  0.9 & 201 & $1.593 \pm 0.3444$ & $2.806 \pm 0.1617$ \\
  1.0 & 159 & $2.557 \pm 0.2284$ & $2.939 \pm 0.1329$\\
  1.1 & 220 & $3.594 \pm 0.7120$ & $3.024 \pm 0.2584$\\
  1.2 & 181 & $9.103 \pm 0.1922$ & $2.648 \pm 0.6395$\\
  1.4 & 228 & $11.26 \pm 1.872$ & $3.124 \pm 0.6200$\\
  1.5 & 227 & $8.427 \pm 2.339$ & $3.574 \pm 0.2804$\\
  1.6 & 260 & $10.17 \pm 1.721$ & $3.654 \pm 0.3155$\\
  1.8 & 200 & $21.37 \pm 7.604$ & $3.983 \pm 0.4113$\\ \hline \hline
\end{tabular}

\caption{Summary of different runs. $\alpha = \frac{\langle \rho v_x v_y - B_x B_y \rangle}{p_0}$ is time averaged from 201 orbits (1 orbit $= T_{\text{orb}} = 2\pi/\Omega$) onwards for $q=0.7$, 126 orbits onwards for $q=0.9,1.0,1.1$, 151 orbits onwards for $q=1.2,1.4,1.5,1.6$, 101 orbits onwards for $q=1.8$ and volume averaged over the disc region. We also quote $\alpha_{\text{mag}} = \langle - B_x B_y \rangle / \langle B^2 \rangle$, which is nearly invariant with shear. Error represents the standard deviation from time averaging.} 
\end{table}

We fix the domain size to be $4H \times 4H \times 8H$ and the resolution to $24$ zones/H. While MRI generated turbulence is known to be sensitive to resolution and domain size (e.g. \cite{2011ApJ...738...84H}, \cite{2014MNRAS.441.1855N}), since we are only interested in studying the effects of shear, we keep all simulation parameters fixed and only vary shear. Table 1 provides a summary of our runs. The third column quotes the value of Shakura-Sunyaev $\alpha = \langle \rho v_x v_y - B_x B_y \rangle/p_0$. Note that for the calculation of $\alpha$ and the turbulent spectra below, we subtract off the background shear ${\bf v} = {\bf v}_{\text{total}} + q \Omega x {\bf e}_y$, where the shear parameter $q \equiv - d \ln \Omega/d \ln r = 3/2$ for Keplerian flows. 

Note that while $\rho, {\bf v}, {\bf B}$ are periodic in $y$ and $z$ and shear periodic in $x$, derived quantities like the electric field, $E = - {\bf v} \times {\bf B}$ are not shear periodic in $x$ because of explicit dependence on $x$ through the background shear, $-q\Omega x$ (for further discussion, see \cite{2014arXiv1405.3991H}).

\subsection{Range of $q$ accessible with shearing box}
Shearing box simulations are known to exhibit epicyclic oscillations for $q<2$ (e.g. \cite{2010ApJS..189..142S}). But for $q>2$, the shearing box equations lead to exponentially growing and decaying solutions for the mean momenta as we discuss below. The equations for the shearing box approximation in the ideal MHD limit are as follows:
\begin{gather}
\frac{\partial \rho}{\partial t} + \nabla \cdot (\rho {\bf v}) = 0, \\
\frac{\partial \rho {\bf v}}{\partial t} + \nabla \cdot (\rho {\bf v} {\bf v} + {\bf T}) = \rho \Omega^2 (2 q {\bf x} - {\bf z}) - 2 {\bf \Omega} \times \rho {\bf v}, \label{eq:NS} \\
\frac{{\partial \bf B}}{\partial t} = \nabla \times ({\bf v} \times {\bf B}),
\end{gather}
where ${\bf B}$ and ${\bf v}$ are the magnetic field and velocity field, $\rho$ is gas density and $p$ is pressure. Here ${\bf T}$ is a stress tensor given by
\begin{equation}
{\bf T} = (p + B^2/2) {\bf I} - {\bf B} {\bf B},
\end{equation}
where ${\bf I}$ is the identity matrix. 

Writing the $x$ and $y$ components separately for the Navier Stokes equation (Eq. \ref{eq:NS}), we get:
\begin{align}
\frac{\partial \rho v_x}{\partial t} + \partial_i (\rho v_i v_x + T_{ix}) &= 2 \Omega \rho \delta v_y, \notag \\
\frac{\partial \rho \delta v_y}{\partial t} + \partial_i (\rho v_i v_y + T_{iy}) &= \Omega \rho v_x (q-2), \notag
\end{align}
where $\delta v_y = v_y + q\Omega x$. Upon volume averaging the above two equations, we get two coupled equations for the volume averaged momenta $\langle \rho v_x\rangle$ and $\langle \rho \delta v_y\rangle $:
\begin{align}
\frac{\partial \langle \rho v_x \rangle}{\partial t} &= 2 \Omega \langle \rho \delta v_y \rangle, \label{epi1} \\
\frac{\partial \langle \rho \delta v_y \rangle}{\partial t} &= \Omega \langle \rho v_x \rangle (q-2). \label{epi2}
\end{align}
which yields the solution that both averaged momentum densities are proportional to $\exp(\pm i \kappa t) $ for $q<2$, or $\sim \exp(\pm \kappa t)$ for $q>2$ where $\kappa^2 = 2\Omega^2 (2-q)$. Note that for the specified periodic boundary conditions, divergence terms vanish owing to the fact that they get converted into surface terms. So this result holds both for hydrodynamics and magnetohydrodynamics since the latter adds only vanishing divergence terms. We want to point out that divergence terms in the energy equation do not vanish for the shearing box (see \cite{1995ApJ...440..742H}, \cite{2014arXiv1405.3991H} for more details).

In the Rayleigh unstable regime $q>2$, fluctuation perturbations are expected to grow exponentially \citep{2012MNRAS.423L..50B}. Using the shearing box approximation, the above solutions show that there will also be growth of the mean momenta if initially the mean momenta are nonzero. This is a physical effect similar to what one would expect for $\kappa^2 < 0$ for a star orbting a galactic center - a perturbation to the orbit of a star with $\kappa^2 < 0$ will make it break away from its orbit and either run away or decay towards the center. However, if initially the mean momenta are set to exactly zero, no growth of the mean momenta is expected. In our shearing box numerical simulations, we still found growth in the mean momenta even when the initial mean momenta are exactly zero, which means that perturbations in the mean arise from numerical noise. Furthermore, the time step in finite volume codes such as ATHENA is inversely proportional to the largest speed in a given cell. So if the mean momenta keep growing, the time step will keep getting smaller and the simulation will eventually crash.

We found only one previous example of a $q=2.1$ simulation using the shearing box approximation (see figure 1 of \cite{1996ApJ...467...76B}). Although these simulations exhibited exponential growth of turbulent kinetic energy, the simulations were stopped at around 2.5 orbits and did not show a sustained saturated turbulent state. We suspect that their run had the same numerical problems that we experienced. We have thus not included the $q>2$ regime in this paper. Even for non-periodic boundaries, the shearing box approximation would still be troublesome in this regime because the spurious growth term will still be there, adding to and likely dominating, any allowed physical fluxes that might not otherwise volume average to zero.

\section{Results of Varying the Shear Parameter}
We study the effect of shear parameter on different physical quantities in this section. Except Fig. \ref{fig:alpha} and \ref{fig:alphavsq} where we report on $\alpha$ averaged over the whole box, we restrict our discussion to the disc region. We do not include statistics from the corona region because density stratification causes large variations on the order of $100\%$, thus making it hard to distinguish between random error and systematic error. Another reason to focus on the disc region is that it is the region where $\beta > 1$, so any theoretical MRI prediction only applies in this region. The corona region is also expected to be more sensitive to the boundary conditions \citep{2014MNRAS.441.1855N}.

Stratified shearing box simulations are known to run into numerical issues (see \cite{2000ApJ...534..398M} for a detailed discussion). Density stratification $(\rho \propto \exp(-z^2))$ leads to very large Alfven speeds away from the midplane. In regions, where $\beta \ll 1$ \citep{2010ApJS..189..142S}, the time step is constrained by the Alfven wave crossing time. Very large Alfven speeds can thus lead to very small steps that eventually halt the simulation.

We tried shear values ranging from $q=0.2$ up to $q=1.8$ but some of the runs crashed because of very small time steps. We only report here on the runs that had a significant number of orbits in the turbulent state. For all of our simulations, we used a density floor of $10^{-4}$. Note that this density floor corresponds to the value of the  average density  at  $|z| \approx 3H$ and thus the exponential fall off in density is truncated and replaced by fixed density above this  $|z|$.  Although raising the density floor can help keep the time steps larger, we did not change our density floor because raising or lowering it can affect  the magnitude of turbulent quantities (see section 3.3 in \cite{2013ApJ...764...66S} for further discussion). 
\subsection{Time history of stresses}

\begin{figure}

  \centering
    \includegraphics[scale=0.5]{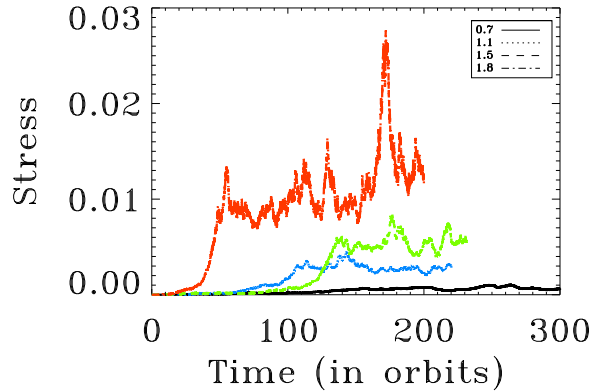}

  \caption{Time history of volume averaged stresses $\alpha = \langle \rho v_x v_y - B_x B_y \rangle/p_0$ for $q=0.7 (\text{black}),1.1 (\text{blue}),1.5 (\text{green}), 1.8 (\text{red})$. } 
\label{fig:alpha}
\end{figure}

We plot the time history of the stress for different runs (Fig. \ref{fig:alpha}) where the stress has been averaged over the whole box. Notice that the $q=0.7$ run does not show noticeable intermittent behaviour while the $q=1.8$ run does. \cite{2008A&A...487....1B} linked the intermittent behaviour to the aspect ratio $L_x/L_y$ for their Keplerian shear runs and found that for $L_x/L_y\ge 1$, the time history of stresses did not exhibit much intermittent behaviour compared to lower aspect ratios. Our domain size (and thus aspect ratio) is fixed but we vary shear and observe this significant change in the stress behaviour with respect to time. This suggests that intermittency is not only related to aspect ratio but also to shear. 

\subsection{Stress and energy}

\begin{figure}

  \centering
    \includegraphics[scale=0.5]{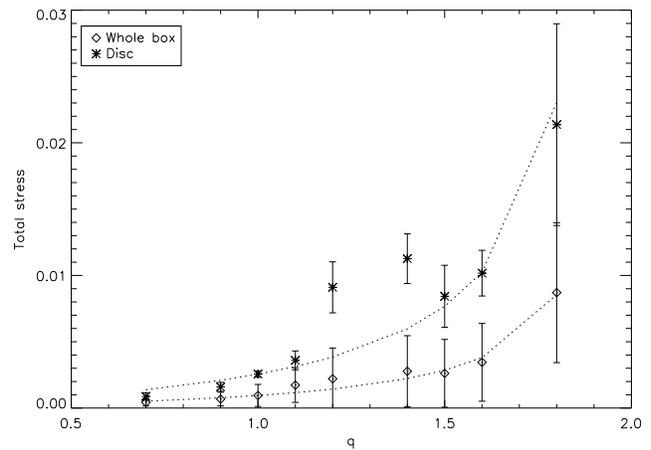}

  \caption{Stress dependence on the shear parameter. The diamonds represent the whole box values while the asterisks represent the disc averaged values. The dotted lines represent the shear to vorticity ratio $q/(2-q)$ fit. There seems to be reasonable agreement between our data for both the disc and volume averaged values and the empirical fit of \citep{1996MNRAS.281L..21A}. Error bars represent the standard deviation due to time averaging.}
\label{fig:alphavsq}
\end{figure}

\begin{figure}

\centering
 \includegraphics[scale=0.5]{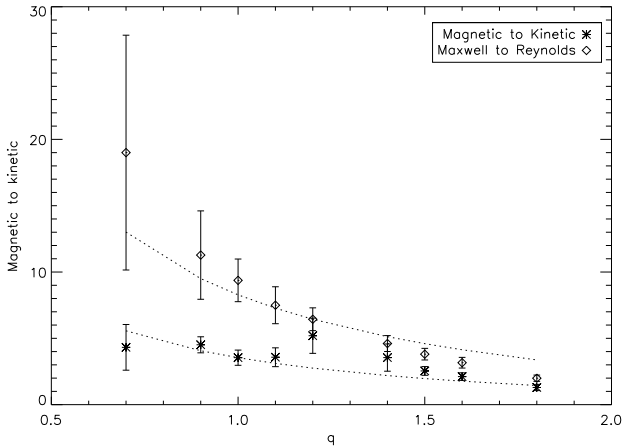}

  \caption{Ratio of Maxwell stress to Reynolds stress and magnetic energy to kinetic energy. The scaling based on the linear regime of the  MRI  $(4-q)/q$ \citep{2006MNRAS.372..183P} is shown by the dashed line.}
\label{fig:magoverkinvsq}
\end{figure}

\begin{figure}

  \centering
    \includegraphics[scale=0.5]{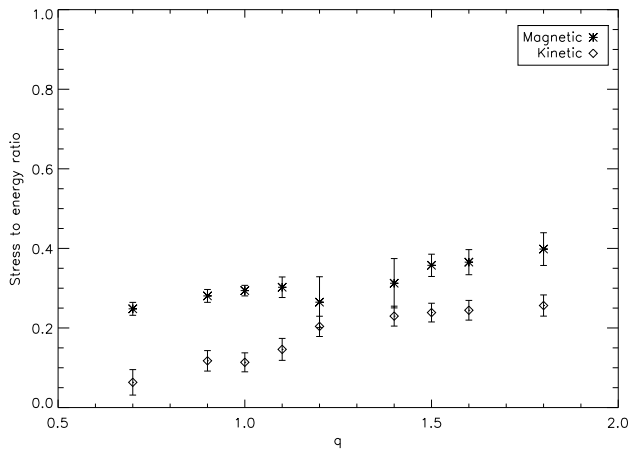}

  \caption{Ratio of Maxwell stress to magnetic energy $\alpha_{\text{mag}}$ and Reynolds stress to kinetic energy. $\alpha_{\text{mag}}$ is a parameter that seems to be an invariant.}
\label{fig:stresstoenergyvsq}
\end{figure}

We plot the total stress against the shear parameter in Fig. \ref{fig:alphavsq}. The dotted line represents the shear to vorticity ratio $q/(2-q)$, something that \cite{1996MNRAS.281L..21A} found the stress is linearly proportional to. The stresses indeed seem to scale linearly with this shear to voriticity ratio. Note that \cite{1996MNRAS.281L..21A} simulations were for the stratified case but with a small vertical domain size $L_z=2H$ whereas we have $L_z=8H$. Thus their calculations were essentially in what we call the disc region. We calculated the volume averaged stresses both in the disc and the whole box and the agreement between the theory and simulations is very good for both regions, which is noteworthy.

The Maxwell to Reynolds stress ratio (Fig. \ref{fig:magoverkinvsq}) gets smaller with shear. The magnetic to kinetic energy ratio seems to follow the same behaviour. This is in agreement with the linear calculations of \cite{2006MNRAS.372..183P} who predicted that the two ratios should depend on the shear parameter as $(4-q)/q$. Note that this theoretical prediction of \cite{2006MNRAS.372..183P} is in agreement with the empirical fit of \cite{1996MNRAS.281L..21A}. 

\subsection{Energy spectra}
\begin{figure}

  \centering
    \includegraphics[scale=0.5]{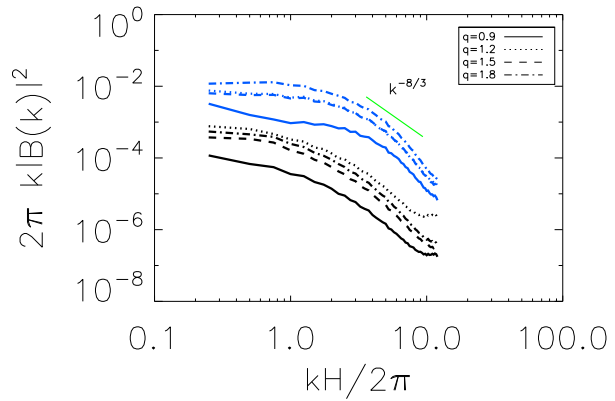}

  \caption{Magnetic energy power spectrum for $q=0.9,1.2,1.5,1.8$ in the disc (blue) and the corona (black) regions. We also show the Kolmogorov fit for circle averaged spectrum $k^{-8/3}$ (green) next to the disc spectra.}
\label{fig:mevsq}
\end{figure}

\begin{figure}

  \centering
    \includegraphics[scale=0.5]{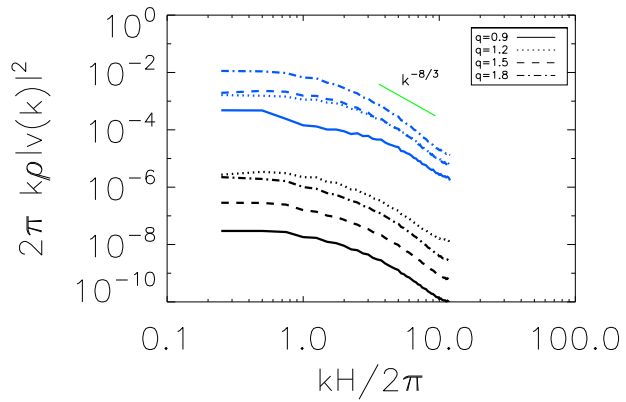}

  \caption{Kinetic energy power spectrum for $q=0.9,1.2,1.5,1.8$. Same color scheme as in Fig. \ref{fig:mevsq}.}
\label{fig:kevsq}
\end{figure}

Using the same method as \cite{2014MNRAS.441.1855N}, we calculate the magnetic (Fig. \ref{fig:mevsq}) and kinetic (Fig. \ref{fig:kevsq}) spectra in the disc and corona region separately. Both energy spectra seem to be relatively insensitive to the shear parameter. We do not see a clear turnover and since we did not use explicit viscosity and resistivity, we cannot specify a physical dissipation scale \citep{2014MNRAS.441.1855N} but it is noteworthy that the spectrum has nearly the same shape for different $q$ values. 

\subsection{Large Scale Field Cycle Period}

\begin{figure*}

  \centering
    \includegraphics[scale=1.0]{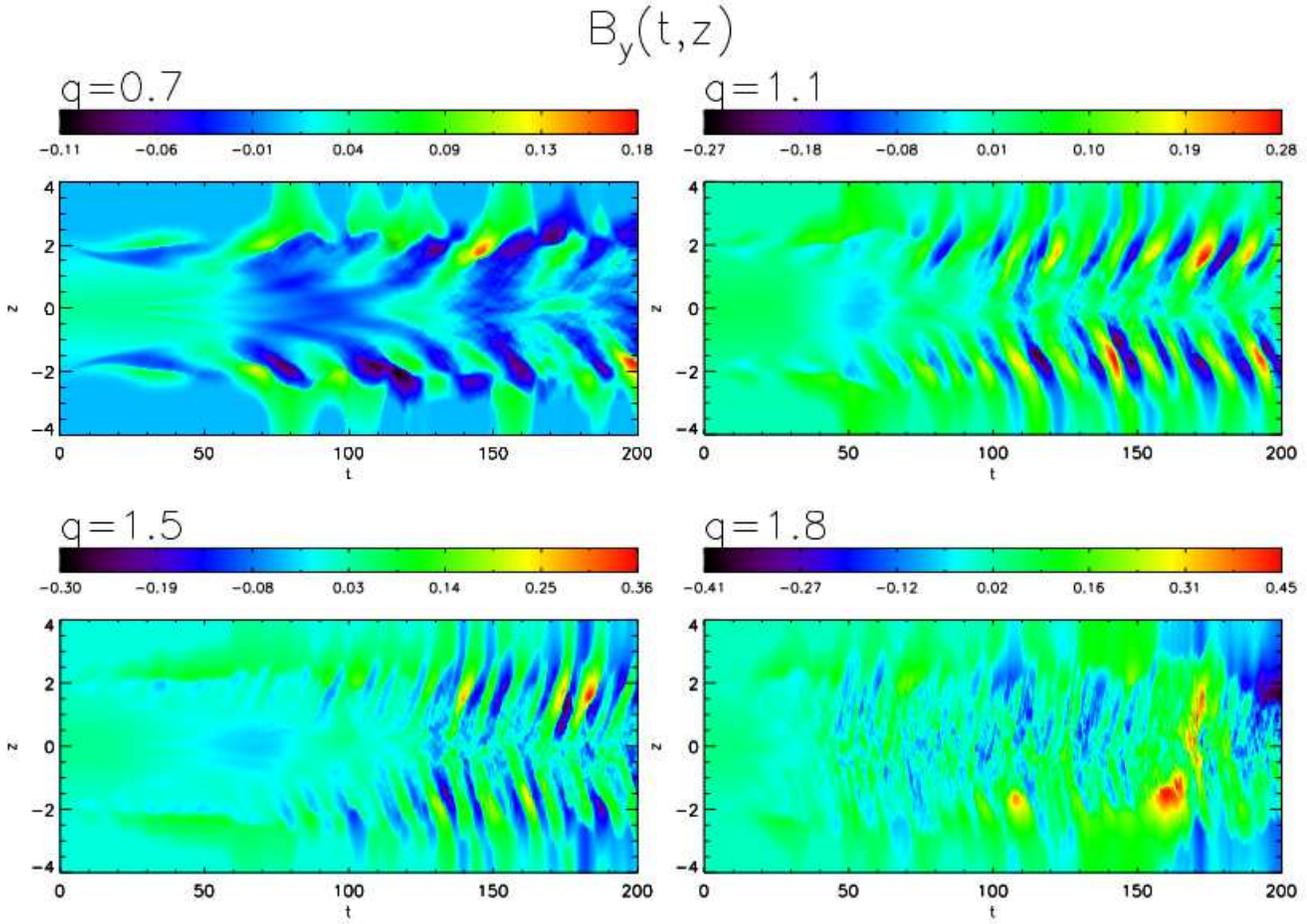}

  \caption{Colour gradient plot of azimuthal field in the $z,t$ plane. $q=0.7$, $q=1.1$ and $q=1.5$ show cycle period of roughly 20, 15 and 10 orbits respectively. It is not easy to estimate the cycle period of $q=1.8$ run but it seems to be shorter than the other 3 runs.}
\label{fig:byvst}
\end{figure*}

We show the colour gradient plot of azimuthal field $B_y(t,z)$ averaged over `x' and `y' in Fig. \ref{fig:byvst} for different shear runs. We find that the cycle period is decreasing with increase in shear. More specifically, $q=0.7$ run seems to have a cycle period of around 20, while $q=1.8$ appears to have a considerably shorter cycle period close to 5 orbits. It is interesting to note that the amplitude of the azimuthal field increases with shear but the period decreases. For Keplerian shear, the cycle period usually turns out to be around $8-10$ orbits for stratified simulations (e.g. \cite{2010ApJ...713...52D}).  So our $q=1.5$ result is consistent with that but we do not know of any earlier study that reported cycle period dependence on shear. The trend of decreased cycle period with increased shear is qualitatively consistent with expectations for generalizing the equations of simple $\alpha-\Omega$ type dynamos \citep{2011ApJ...728..130G}.

\subsection{Tilt angle and $\alpha_{\text{mag}}$}

\begin{figure*}

  \centering
    \includegraphics[scale=1.0]{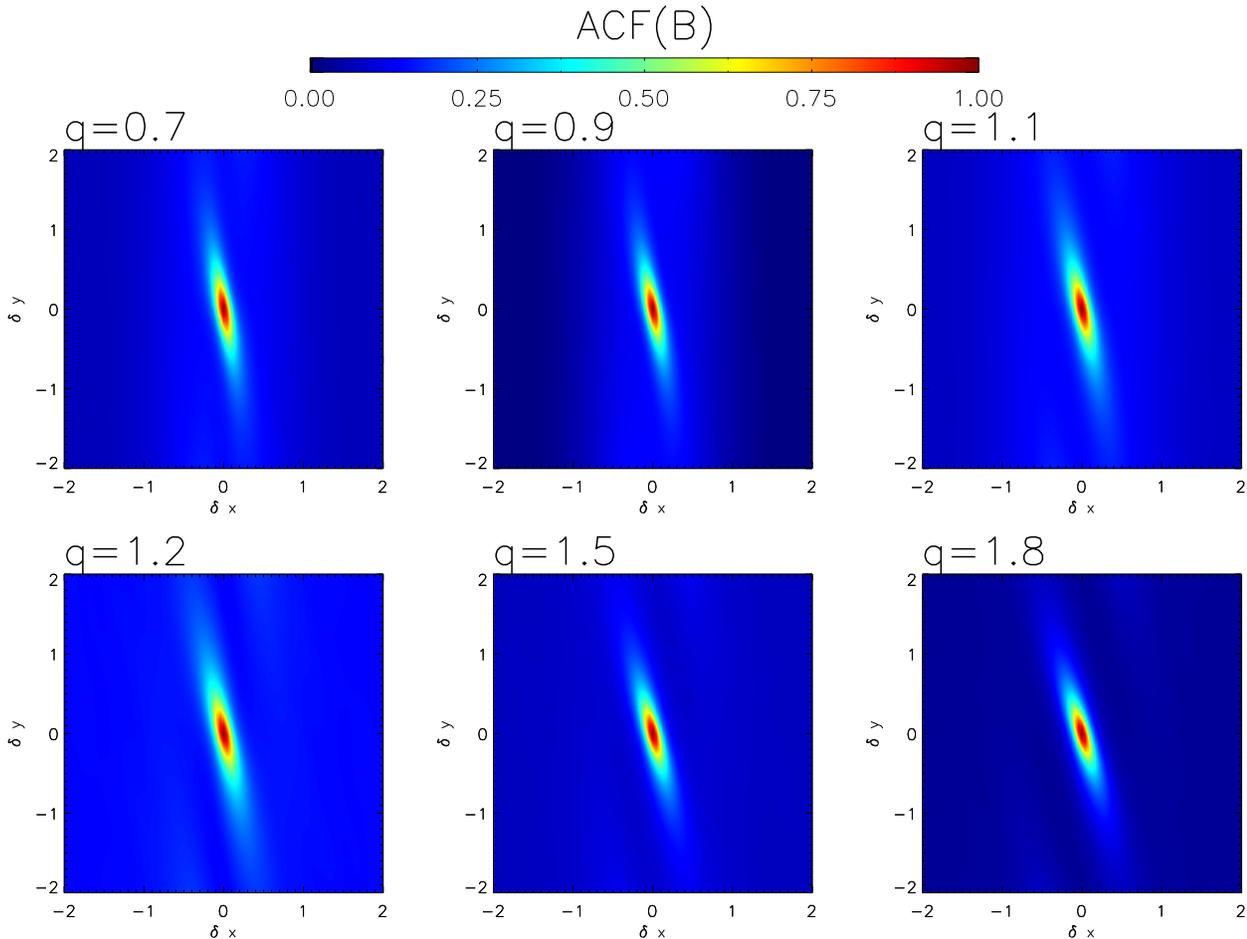}

  \caption{Autocorrelation of the magnetic field, $ACF (B)$ as defined in Eq. \ref{eq:ACF} for 6 different shear runs $q=0.7,0.9,1.1,1.2,1.5,1.8$. The horizontal axis is the lag in x-direction and the vertical axis is the lag in y-direction. The contours are tilted at an angle between $10-15$ degrees with respect to the y-axis and this tilt angle seems to be roughly independent of the shear.}
\label{fig:contour2D}
\end{figure*}

The Maxwell stress to magnetic energy ratio, $\alpha_{\text{mag}}$, (Fig. \ref{fig:stresstoenergyvsq}) seems to be roughly unaffected by the shear parameter. This can also be seen from the autocorrelation plots of the magnetic field. Following the convention used by \cite{2009ApJ...694.1010G} and \cite{2012MNRAS.422.2685S}, we define the autocorrelation of the magnetic field component `i' ($i=x,y,z$):
\begin{equation}
\text{ACF} (B_i({\bf \delta x}) ) = \left\langle \frac{\int B_i ({\bf x + \delta x},t) B_i({\bf x}, t) d^3 {\bf x}} {\int B_i^2({\bf x}, t) d^3 {\bf x}} \right\rangle
\label{eq:ACF}
\end{equation}

Note that ACF($B_i$) is normalized to its maximum value at zero lag ($\delta x = \delta y = \delta z = 0$). We subtract off the pertinent mean quantities (i.e. mean magnetic field for the magnetic field ACF) as in \cite{2009ApJ...694.1010G} and calculate the total ACF associated with fluctuations as ACF($B$) = ACF($B_x$) + ACF($B_y$) + ACF($B_z$). 

Angled brackets represent time averaging over several orbits in saturated state. We do the spatial integration over all x and y but in the vertical direction we only consider the disc region (defined by the region where $\beta > 1$). The result is shown in Fig. \ref{fig:contour2D}. The tilt angle with respect to the y-axis is proportional to $\alpha_{\text{mag}}$ \citep{2009ApJ...694.1010G} and a visual inspection shows that it is roughly invariant with respect to shear. This is consistent with our calculation of $\alpha_{\text{mag}}$ in table 1 which also varies very little with shear.

This insensitivity of $\alpha_{\text{mag}}$ to $q$ is a particularly noteworthy result as $\alpha_{\text{mag}}$ is known be an invariant quantity from previous Keplerian shearing box simulations (e.g. \cite{2008NewA...13..244B}, \cite{2011ApJ...738...84H}), i.e. it does not depend on resolution, domain size, initial field strength, dissipation coefficients, etc. While previous studies did not consider the effect of changing $q$, the near constancy of ACF($B$) with $q$ highlights a further robustness of the near constancy of $\alpha_{\text{mag}}$. 

We find that it can be explained by the confluence of two competing effects as shear is increased. The induction equation implies that, from an initial radial field $B_x$, an azimuthal component of the field will be amplified by shear scaling roughly as
\begin{equation}
B_y \sim -q\Omega\tau B_x,
\label{bygrowth}
\end{equation}
where $\tau$ is the correlation time scale. Roughly $\alpha_{\text{mag}} \sim B_x B_y/B^2 \sim B_x/B_y$. So if the the correlation time $\tau$ goes as $1/q$, then this ratio and thus $\alpha_{\text{mag}}$ will be roughly constant with shear. This is indeed the case as we discuss in the next section.

\subsection{Correlation time}

\begin{figure}
\centering
  \includegraphics[scale=0.5]{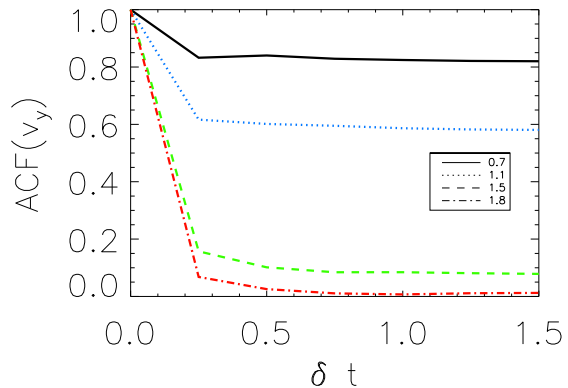}
  \caption{Auto-correlation plot of azimuthal velocity $v_y$ for $\beta = 1600$ runs sampled at $0.25$ orbits. As can be seen, because of smaller sampling frequency, we cannot observe an exponential decay in the correlation and thus it is not easy to calculate the correlation time.}
  \label{fig:corrtime1600}
\end{figure}
\begin{figure}
  \centering
  \includegraphics[scale=0.5]{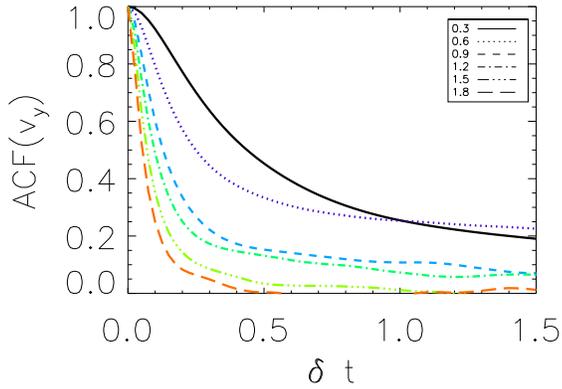}
  \caption{Auto-correlation plot of azimuthal velocity $v_y$ for the new $\beta = 100$ runs sampled at $0.025$ orbits. Higher sampling rate allows us to see the exponential decay clearly so that we can calculate the correlation time by using an exponential fit.}
  \label{fig:corrtime100}
\end{figure}
\begin{figure}
  \centering
  \includegraphics[scale=0.5]{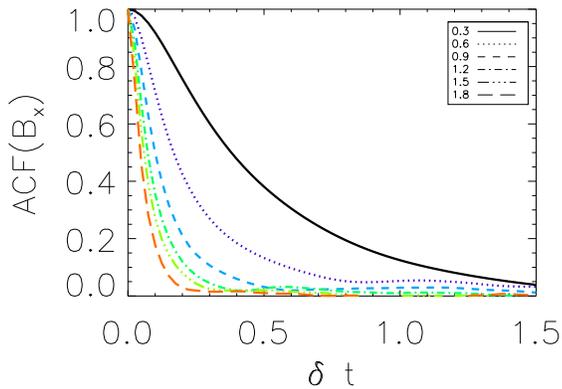}
  \caption{Same as Fig. \ref{fig:corrtime100} but for the radial magnetic field $B_x$.} 
  \label{fig:corrtime100bx}
\end{figure}
\begin{figure}
  \centering
  \includegraphics[scale=0.5]{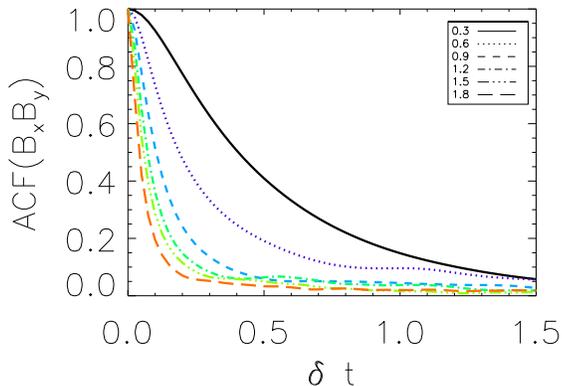}
  \caption{Same as Fig. \ref{fig:corrtime100} but for Maxwell stress $B_xB_y$.}
  \label{fig:corrtime100mxy}
\end{figure}
\begin{figure}
  \centering
  \includegraphics[scale=0.5]{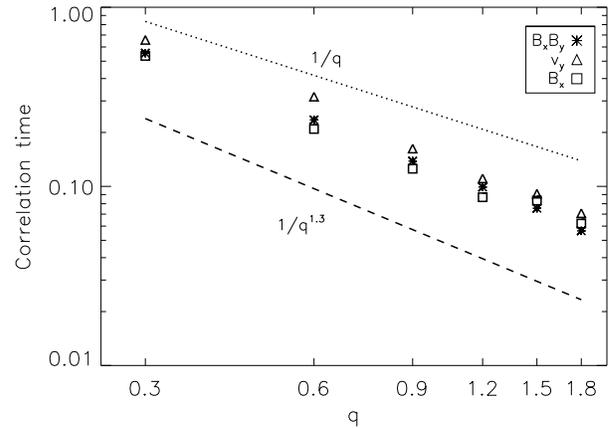}
  \caption{Log-log plot for $\beta = 100$ runs showing correlation time for $v_y$ (diamonds), $B_xB_y$ (asterisks) and $B_x$ (squares) calculated by an exponential fit. The dotted line shows the $1/q$ fit, while the dashed line the $1/q^{1.3}$ fit.}
  \label{fig:oneoverq}
\end{figure}

To get an estimate of the time scale on which the azimuthal field $B_y$ is generated from the radial field $B_x$, we calculated the stretching time scale, which, from the induction equation for $B_y$ depends on the correlation time of $B_x$ in the following sense: The correlation time $\tau$ in Eq. (\ref{bygrowth}) arises from the induction equation $\partial_t B_y =  -q B_x \Omega$, if $B_x$ is nearly constant over a time $\tau$. $B_x$ can change by advection, decay, etc. Thus we are interested in the autocorrelation time of $B_x$.

If we consider the equation for $B_y^2$, then the autocorrelation of $B_x B_y$ would be germane
in the same context. Since we are interested in why $\alpha_{\text{mag}}$ is invariant with shear, this correlation time is particularly relevant. We plot these two correlation times as a function of $q$, and for comparison also add the autocorrelation time of $v_y$, which is the largest component of the velocity. 
 
Specifically, we define the auto-correlation in time of a physical quantity `f' the following way:
\begin{equation}
\text{ACF} (f(\delta t) ) = \left\langle \frac{\int f ({\bf x},t+\delta t) f({\bf x}, t) dt} {\int f^2({\bf x}, t) dt} \right\rangle,
\label{eq:ACFt}
\end{equation}
where the angle brackets represent volume averaging over all $x$, $y$ but $z=0H$ to $z=1H$. Time integration is done over several orbits in the turbulent saturated state. 

We calculated the autocorrelation time scales for the runs reported earlier (Fig. \ref{fig:corrtime1600}) and found that the correlation times decrease with increase in shear but the sampling interval of only $0.25 T_{\text{orb}}$ was too low to identify the dependence on $q$. To better quantify the relation of correlation times with the shear parameter, we had to run another set of simulations with a higher sampling frequency. We used the same parameters as our earlier simulations except for a smaller domain size $1H \times 2H \times 4H$, a resolution of 32 zones/H and smaller initial plasma beta (ratio of thermal to magnetic pressure) $\beta=100$. We did this because we sampled the data at $0.025 T_{\text{orb}}$ for these runs and so the data we collected is enormous. We ran 6 different shear values $q=0.3,0.6,0.9,1.2,1.5,1.8$.

We calculated the autocorrelation of $B_x$, $B_xB_y$ and $v_y$, and the results are shown in Fig. \ref{fig:corrtime100}, \ref{fig:corrtime100bx} and \ref{fig:corrtime100mxy}. By using an exponential fit, we found that the correlation time dependence on `q' resembles $\tau \sim 1/q^{w}$ where $1\le w\le 1.3$  as seen  in Fig. \ref{fig:oneoverq}. This supports our earlier claim that the near invariance of $\alpha_{\text{mag}}$ with shear is because the correlation time is roughly inversely proportional to the shear parameter. 

\section{Conclusion}
We have explored the sensitivity of the MRI on the shear parameter for $q<2$, using ideal MHD stratified shearing box simulations of modest resolution of 24 zones/H with domain size $4H \times 4H \times 8H$. We found that certain physical quantities depend on the shear parameter while others do not:
\begin{enumerate}
\item Turbulent stresses always increase with the shear parameter and they scale linearly with the shear to vorticity ratio ($q/(2-q)$), in agreement with earlier calculations of \cite{1996MNRAS.281L..21A}.
\item The ratio of Maxwell to Reynolds stress and magnetic to kinetic energy seems to closely follow the scaling $(4-q)/q$ that \cite{2006MNRAS.372..183P} predicted using linear MRI analysis. 
\item The cycle period of the azimuthal field depends on shear and it decreases with increasing shear, while the amplitude of the azimuthal field increases with shear.
\item The shape of the turbulent kinetic and magnetic energy spectrum do not depend strongly on the shear parameter. 
\item $\alpha_{\text{mag}}$ ($=\langle -B_xB_y \rangle/\langle B^2 \rangle$) is nearly invariant with respect to the shear parameter. Using a simple argument that $B_y \sim - q \Omega \tau B_x$ where $\tau$ is a turbulent correlation time, we showed that the azimuthal field amplification is independent of the shear parameter because the correlation time roughly varies as $1/q$. 
\end{enumerate}

We emphasize that in a shearing box simulation, there is no feedback on the background shear whereas in actual accretion discs, one expects the rotation profile to be modified by thermal pressure and external forces. Nevertheless, comparative analysis of shearing box simulations with different shear values--treated as a set of mutually self-consistent numerical experiments--can help us improve our understanding of the physics of  MRI driven turbulence. Building on \cite{1996MNRAS.281L..21A} and \cite{2006MNRAS.372..183P}, for example, the study can help to better inform mean field models beyond the Shakura-Sunayev which does not capture the correct $q$ dependence of the stress.

\bibliography{general}
\bibliographystyle{mn2e}

\section*{Acknowledgments}
We thank U. Torkelsson for constructive comments, and R. Penna for discussions. FN acknowledges Horton Fellowship from the Laboratory for Laser Energetics at U. Rochester and we acknowledge support from NSF grant AST-1109285. EB acknowledges support from the Simons Foundation and the IBM-Einstein Fellowship fund at IAS. We acknowledge the Center for Integrated Research Computing at the University of Rochester for providing computational resources.
\bsp

\label{lastpage}

\end{document}